\documentstyle[12pt,graphicx]{article}
\begin{document}

\def\ds{\displaystyle}
\def\beq{\begin{equation}}
\def\eeq{\end{equation}}
\def\bea{\begin{eqnarray}}
\def\eea{\end{eqnarray}}
\def\beeq{\begin{eqnarray}}
\def\eeeq{\end{eqnarray}}
\def\ve{\vert}
\def\vel{\left|}
\def\ver{\right|}
\def\nnb{\nonumber}
\def\ga{\left(}
\def\dr{\right)}
\def\aga{\left\{}
\def\adr{\right\}}
\def\lla{\left<}
\def\rra{\right>}
\def\rar{\rightarrow}
\def\nnb{\nonumber}
\def\la{\langle}
\def\ra{\rangle}
\def\ba{\begin{array}}
\def\ea{\end{array}}
\def\tr{\mbox{Tr}}
\def\ssp{{\Sigma^{*+}}}
\def\sso{{\Sigma^{*0}}}
\def\ssm{{\Sigma^{*-}}}
\def\xis0{{\Xi^{*0}}}
\def\xism{{\Xi^{*-}}}
\def\qs{\la \bar s s \ra}
\def\qu{\la \bar u u \ra}
\def\qd{\la \bar d d \ra}
\def\qq{\la \bar q q \ra}
\def\gGgG{\la g^2 G^2 \ra}
\def\q{\gamma_5 \not\!q}
\def\x{\gamma_5 \not\!x}
\def\g5{\gamma_5}
\def\sb{S_Q^{cf}}
\def\sd{S_d^{be}}
\def\su{S_u^{ad}}
\def\ss{S_s^{??}}
\def\sbp{{S}_Q^{'cf}}
\def\sdp{{S}_d^{'be}}
\def\sup{{S}_u^{'ad}}
\def\ssp{{S}_s^{'??}}
\def\sig{\sigma_{\mu \nu} \gamma_5 p^\mu q^\nu}
\def\fo{f_0(\frac{s_0}{M^2})}
\def\ffi{f_1(\frac{s_0}{M^2})}
\def\fii{f_2(\frac{s_0}{M^2})}
\def\O{{\cal O}}
\def\sl{{\Sigma^0 \Lambda}}
\def\es{\!\!\! &=& \!\!\!}
\def\ar{&+& \!\!\!}
\def\ek{&-& \!\!\!}
\def\cp{&\times& \!\!\!}
\def\se{\!\!\! &\simeq& \!\!\!}
\def\kpm{&\pm& \!\!\!}
\def\kmp{&\mp& \!\!\!}

% .........................................................

\def\simlt{\stackrel{<}{{}_\sim}}
\def\simgt{\stackrel{>}{{}_\sim}}

% .........................................................

\title{
         {\Large
                 {\bf
  Radiative $ \rho\rar \eta\gamma $ decay in light cone
 QCD
                 }
         }
      }

\author{\vspace{1cm}\\
{\small T. M. Aliev$^a$ \thanks {e-mail: taliev@metu.edu.tr}\,\,,
{\. I} Kan{\i}k$^a$ \thanks {e-mail: e114288@metu.edu.tr}\,\,, A.
\"{O}zpineci$^b$ \thanks {e-mail: ozpineci@ictp.trieste.it}\,\,,}
\\
{\small a Physics Department, Middle East Technical University,
06531 Ankara, Turkey}\\
{\small b  The Abdus Salam International Center for Theoretical Physics,
I-34100, Trieste, Italy} }
\date{}

\begin{titlepage}
\maketitle
\thispagestyle{empty}

\begin{abstract}
The coupling constant of $ \rho\rar \eta\gamma $ decay is
calculated in the framework of light cone QCD sum rules. A
comparison of our prediction on the coupling constant with the
result obtained from analysis of the experimental data is
performed.
\end{abstract}
%\vspace{1cm}
~~~PACS numbers:~~11.55.Hx, 13.40.Hq.
\end{titlepage}

\section{Introduction}
    Radiative transition between vector(V) and pseudoscalar(P)
mesons represent an important source of information on low energy
hadron physics. These transitions are governed by magnetic dipole
(M1) radiation of the photon and had played one of the central
roles for checking the predictions of quark model and SU(3)
symmetry, as well as it was very useful in the determination of the magnetic dipole moment of $N^*(1535)$
in $\gamma N \rightarrow \eta N$ process \cite{R20}. Recently the theoretical activity on the VP$\gamma$
magnetic dipole transition have increased(see \cite{R1} and
references therein), in particular,
due to the fact that the analysis of the radiative $V \rar P\gamma
$ decay with $\eta$ and $\eta'$ mesons in final state can provide
insights to the long standing issue of the $\eta$ and $\eta'$
mixing (for review see \cite{R2} and references therein). The
usual parametrization of $\eta - \eta'$ mixing in the octet--singlet basis, which will be used in this work, is as follows:
the current-particle matrix elements are defined as \bea \la 0 |
J_{5\mu}^\alpha | P(p) \ra = -i f_P^\alpha p_\mu
(\alpha=8;0;P=\eta,\eta' \label{eq1}) \eea where $ J_{5\mu}^8$,
the $SU(3)_F$ octet axial vector current, is given by \bea
J_{5\mu}^8=\frac{1}{\sqrt{6}}(\overline{u}\gamma_\mu\gamma_5
u+\overline{d}\gamma_\mu\gamma_5 d-2\overline{s}\gamma_\mu\gamma_5
s) \label{eq2} \eea
 and $ J_{5\mu}^0$, the $SU(3)_F$ singlet current is given by:
\bea
J_{5\mu}^0=\frac{1}{\sqrt{3}}(\overline{u}\gamma_\mu\gamma_5
u+\overline{d}\gamma_\mu\gamma_5 d+\overline{s}\gamma_\mu\gamma_5
s)
\label{eq3}
\eea

Two mixing angles $\theta_8$ and $\theta_0$ are required in order
to consistently describe mixing \cite{R3}. Accordingly the
couplings in Eq.(\ref{eq1}) can be defined as follows
\bea
f_\eta^8 = f_8 cos\theta_8  \; &,& \;\;\;   f_\eta^0 = - f_0 sin\theta_0 \\
f_{\eta'}^8 = f_8 sin\theta_8 \; &,& \;\;\; f_{\eta'}^0 = f_0 cos\theta_0
\label{eq4}
\eea

%The recent analysis shows that both mixing angles are about $-10^\circ$ .
 Alternatively, two independent
axial vector currents with distinct flavors can be considered \bea
J_{5\mu}^q &=& \frac{1}{\sqrt{2}}(\overline{u}\gamma_\mu\gamma_5
u+\overline{d}\gamma_\mu\gamma_5 d) \nnb \\ J_{5\mu}^s &=&
\overline{s}\gamma_\mu\gamma_5s \label{eq5} \eea and the couplings
of these currents with $\eta$ and $\eta'$ mesons are defined
similarly to Eq.(\ref{eq1}) : \bea
f_\eta^q = f_q cos\varphi_q \; &,& \;\;\;  f_{\eta'}^q = f_q sin\varphi_q \\
f_\eta^s = -f_s sin\varphi_s  \; &,& \;\;\; f_{\eta'}^s =f_s cos\varphi_s
\label{eq6}
\eea

As we see that in each basis there are two angles and in \cite{R2}
it was shown that, to a very good accuracy, the mixing can described
in terms of single angle $\varphi$ since $ |\varphi_s -
\varphi_q|/|\varphi_s + \varphi_q| << 1$ which is also  confirmed by a QCD
sum rules calculation \cite{R4}.

In present work we will follow to the first approach, neglecting
the mixing angles $\theta_0$ and $\theta_8$ due to their smallness
and calculate the coupling constant of $\rho\rightarrow\eta\gamma$
decay in framework of light cone QCD sum rules (more about of
light cone QCD sum rules and its applications can be found in
\cite{R5} and \cite{R6}).

The paper is organized as follows: In section 2 we derive light
cone QCD sum rules for $\rho\rightarrow\eta\gamma$ decay constant,
section 3 is devoted to the numerical analysis of the sum rules and contain
our conclusions.

\section{Light Cone QCD Sum Rules for $\rho\rightarrow\eta\gamma$ coupling constant}

In this section we calculate the coupling constant of
$\rho\rightarrow\eta\gamma$ decay using light cone QCD sum rules
method. In order to calculate this coupling constant we consider
the following two point correlation function
 \bea \Pi_{\mu\nu} = i
\int d^4 x e^{i p_2 x} \ \la 0 | T\left\{ J_\nu^\eta(x)
J_\mu^\rho(0)\right\} | 0 \ra_\gamma \label{eq7} \eea where
$\gamma$ denotes the  external electromagnetic field, and
$J_\nu^\eta$ and $J_\mu^\rho$ are the interpolating currents with
$\eta$ and $\rho$ meson quantum numbers. Here we would make the
following remark. As we already noted that both of the mixing
angles in Eq. (\ref{eq4}) are small, in the following discussion
we will neglect the mixing, i.e we take $J_\mu^\eta \equiv
J_{5\mu}^8$.

At the phenomenological level the Eq.(\ref{eq7}) can be expressed
as:
\bea
\Pi_{\mu\nu} = \sum \frac{\la 0 | J_\nu^\eta | \eta(p_2) \ra
\la \eta(p_2) |\rho(p_1)\ra_\gamma \la \rho(p_1) | J_\mu^\rho | 0
\ra}{(p_2^2-m_\eta^2)(p_1^2-m_\rho^2)}
\label{eq8}
\eea
where $p_1 = p_2 + q $ and q is the photon momentum. The matrix
elements entering Eq.(\ref{eq8}) are defined as
\bea
\la 0 | J_\mu^\rho | \rho \ra &=& m_\rho f_\rho \varepsilon_\mu^\rho \label{eq9} \\
\la 0 |J_{5 \nu}^8| \eta(p_2)\ra &=& -i f_\eta^8 p_{2\nu}
\label{eq10} \eea
%where
%for simplicity we use $\overline{q} \gamma_\nu \gamma_5 q$
%to denote $J_{5\mu}^8=\frac{1}{\sqrt{6}}(\overline{u}\gamma_\mu\gamma_5
%u+\overline{d}\gamma_\mu\gamma_5 d-2\overline{s}\gamma_\mu\gamma_5
%s)$ and also introduce $F_\eta = \frac {f_\eta} {\sqrt{6}}$ ;
%since considered problem only u and d quarks active quarks so in
%right side of Eq.(\ref{eq10}) will appear $F_\eta^u$ or
%$F_\eta^d$.
The remaining matrix element $\la \eta(p_2)| \rho(p_1)\ra_\gamma$
which describes the M1 transition, can be parameterized as, \bea
\la \eta(p_2)| \rho(p_1)\ra_\gamma = e
\epsilon_{\mu\nu\alpha\beta}\varepsilon_\mu^\rho p_{1\nu}
\varepsilon_\alpha^\gamma q_\beta F(q^2) \label{eq11} \eea where
$\varepsilon^\gamma$ is the photon polarization vector. Since the
photon is real, we need the value of $F(q^2)$ only at the point
$q^2=0$. We can use an alternative parametrization for the
$\rho\eta\gamma$ vertex: \bea L_{int} = - \frac{e}{m_\rho}
g_{\rho\eta\gamma}\epsilon_{\mu\nu\alpha\beta}
(\partial_\mu\rho_\nu-\partial_\nu\rho_\mu)(\partial_\alpha
A_\beta - \partial_\beta A_\alpha) \label{eq12} \eea

Comparing Eqns. (\ref{eq11}) and (\ref{eq12}) we see that
\bea
F(q^2=0)\equiv \frac{g_{\rho\eta\gamma}}{m_\rho}
\label{eq13}
\eea

Using Eqns. (\ref{eq8}) -- (\ref{eq12}), for the physical part of
the sum rules we get \bea \Pi_{\mu\nu}^{ph} =
\frac{g_{\rho\eta\gamma}f_\rho f_\eta^8
p_{2\nu}\epsilon_{\mu\sigma\alpha\beta} p_{1\sigma}
\varepsilon_\alpha^\gamma q_\beta}
{(p_2^2-m_\eta^2)(p_1^2-m_\rho^2)} \label{eq14} \eea

Our next task is the calculation of correlator Eq.(\ref{eq1}) from the
QCD side. The correlator receives both perturbative and non-perturbative
contributions. In calculation of the non-perturbative contributions by the OPE on the
light cone one needs to know the matrix elements of nonlocal
operators between vacuum and the photon states;i.e.
$\la \gamma(q)|\bar q \Gamma_i q |0 \ra$ where $\Gamma_i$ is an arbitrary
Dirac matrix. These matrix elements can be expressed in terms of photon
wave functions with definite twist. In calculations we neglect
twist 3 three particle photon wave functions since their
contributions are small. Twist two and twist four photon wave functions are defined
as \cite{R7,R8,R9}:
\bea
\la \gamma(q) | \bar q \sigma_{\alpha\beta}q |0\ra &=&
i e_q \la\bar q q\ra \int_0^1 du e^{iuqx} \{(\varepsilon_\alpha q_\beta - \varepsilon_\beta
q_\alpha)[\chi\varphi(u)+x^2 \left[g_1(u)-g_2(u)\right]] + \nnb \\
&&[qx(\varepsilon_\alpha x_\beta - \varepsilon_\beta x_\alpha)+
\varepsilon x (x_\alpha q_\beta - q_\alpha x_\beta)]g_2 (u)\}
\label{eq16} \\
\la \gamma(q) | \bar q \gamma_\alpha \gamma_5q |0\ra  &=&
\frac{f}{4} e_q \epsilon_{\mu\alpha\beta\rho}
\varepsilon^\alpha q^\beta x^\rho \int_0^1 du e^{iuqx}\psi(u)
\label{eq17}
\eea
where
for simplicity we use $\overline{q} \Gamma q$
to denote $J_{5\mu}^8=\frac{1}{\sqrt{6}}(\overline{u}\Gamma u+\overline{d} \Gamma d-2
\overline{s}\Gamma s)$.% and also introduce $F_\eta = \frac {f_\eta} {\sqrt{6}}$.

In Eqs.(\ref{eq16}) and (\ref{eq17}) $e_q$ is the corresponding
quark charge, $\chi$ is the magnetic susceptibility, $\varphi(u)$, and $\psi(u)$ are the leading
twist two and $g_1(u)$, and $g_2(u)$ are the twist four photon wave
functions.

After standard calculations we get the following expression for
correlator from QCD side in the coordinate representation: \bea
\Pi_{\mu\nu} = && \frac{1}{\sqrt{12}}  \frac{i}{16}(e_u - e_d)
\int d^4 x e^{i p_2 x} \{[6(\varepsilon
x)\epsilon_{\alpha\beta\nu\mu} q_\alpha x_\beta -
6(qx)\epsilon_{\alpha\beta\nu\mu}\varepsilon_\alpha x_\beta +
6x^2\epsilon_{\alpha\beta\nu\mu}\varepsilon_\alpha q_\beta \nnb \\
&& -12 x_\nu \epsilon_{\alpha\beta\rho\mu} \varepsilon_\alpha
q_\beta x_\rho]+f\pi^2 x^2
[x_\nu\epsilon_{\alpha\beta\rho\mu}\varepsilon_\alpha q_\beta
x_\rho+ \nnb \\ &&
x_\mu\epsilon_{\alpha\beta\rho\nu}\varepsilon_\alpha q_\beta
x_\rho]\int_0^1 du e^{iuqx}\psi(u)\}/\pi^4 x^6 \label{eq18} \eea

The sum rules for $g_{\rho\eta\gamma}$ are obtained by equating the
phenomenological and theoretical parts (in Eq.(\ref{eq18}) it is
necessary to perform Fourier transformation first) of the
correlator.

Performing double Borel transformation on variables $p_2^2 = p^2$
and $p_1^2 = (p+q)^2$ on both sides of the correlation function in
order to suppress the contributions of the continuum and higher
states (procedure of subtraction in light cone sum rules one can
find in \cite{R10,R11}), and also to remove the subtraction terms in the dispersion relation,
we obtain the following sum rules for
$g_{\rho\eta\gamma}$ coupling constant
\bea
g_{\rho\eta\gamma}=
\frac {e^{[m_\rho^2/M_1^2 + m_\eta^2/M_2^2]}(e_u-e_d)}{2 \sqrt{3}
f_\eta^u f_\rho} (1- u_0) \{ \frac {3}{2\pi^2} M^2 E_0(s_0/M^2) +
f \psi(u_0) \}
\label{eq19}
\eea
where
$E_0(s_0/M^2)=1-e^{-s_0/M^2}$ is the function used to subtract
continuum, $s_0$ is the continuum and
\bea
u_0= \frac{M_1^2}{M_1^2+M_2^2}  , M^2=\frac{M_1^2 M_2^2}{M_1^2+M_2^2}
\label{eq15}
\eea
where $M_1^2$ and $M_2^2$  are the Borel mass
parameters in $\rho$ and $\eta$ channels.
Note that in Eq.(\ref{eq15}) we take into account
$F_\eta^u=F_\eta^d$.  The masses of $\eta$ meson
is close to the $\rho$ meson mass. For this reason, it is natural to set
$M_1^2=M_2^2=2M^2$ from which it follows that $u_0 = 1/2$.

\section{Numerical Analysis}

In this section we present our numerical result on
$g_{\rho\eta\gamma}$ coupling constant. From sum rules
Eq.(\ref{eq19}) we see that for estimating $g_{\rho\eta\gamma}$
coupling constant first of all one needs to know the photon wave
function $\psi(u)$. It was shown in \cite{R8,R9} that the photon
wave function do not deviate remarkably from its asymptotic form
which is given by $\psi(u)=1$ \cite{R8,R7}. The values of the
other constants appearing in the sum rules are: $m_\rho =
0.77~GeV$, $m_\eta=0.55~GeV$, $f= 0.028 GeV^2$. Leptonic decay
constant of $\rho$ meson $f_\rho = 0.15~GeV$ follows from
experimental result of the $\rho \rightarrow e^+ e^- $ decay ,
$\Gamma(\rho \rightarrow e^+ e^-)= (6,85 \pm 0,11)KeV $\cite{R13}.
More recent analysis shows that the coupling of $\eta$ meson with
the octet axial vector current is $f_\eta^8 = 0.159~GeV$ \cite{R2}
and this result we will use in our analysis.

In Fig. 1 we present the dependence of the coupling constant
$g_{\rho\eta\gamma}$ on the Borel parameter $ M^2 $ at three
different values of the continuum threshold: $s_0= 1.4~GeV^2,~
1.6~ GeV^2,~ 1.8~GeV^2 $. Since the Borel mass $ M^2 $ is an
auxiliary parameter and the physical quantities should not depend
on it, we must look for the region where $g_{\rho\eta\gamma}$ is
practically independent of $M^2$. We obtain that this condition is
satisfied when $1~GeV^2 \leq M^2 \leq 1.4~GeV^2$. From this figure
we also obtained that the variation of $s_0$ from $s_0= 1.4~GeV^2$
to $s_0= 1.8~GeV^2$ causes a change on the result on
$g_{\rho\eta\gamma}$ of about $10\%$. Therefore one can say that
the result $g_{\rho\eta\gamma}$ is insensitive to $s_0$ and $M^2$.
Our final prediction on the coupling constant is \bea
g_{\rho\eta\gamma}=(1.4 \pm 0.2) \label{eq20} \eea where the
error is attributed to the variation of $s_0$, $M^2$ and neglected
twist three photon wave functions.
%If we take into account the mixing
%angle $cos\theta_8 \cong cos(-10^\circ)\approx 0.78$ , we must
%divide the right side of Eq.(\ref{eq19}) by $0.78$, and we get
%\bea
%g_{\rho\eta\gamma}=(1.41 \pm 0.13)
%\eea

At the end we would like to compare our prediction on
$g_{\rho\eta\gamma}$ with experimental result. The decay width of
the $\rho\rightarrow\eta\gamma$ decay is given by \bea
\Gamma(\rho\rightarrow\eta\gamma)=\frac{\alpha
g^2_{\rho\eta\gamma}}{24}m_\rho (1-m_\eta^2/m_\rho^2)^3
\label{eq21} \eea The experimental value is
$\Gamma(\rho^0\rightarrow\eta\gamma) = (57 \pm 10)~
KeV$\cite{R13}. Using this value of
$\Gamma(\rho^0\rightarrow\eta\gamma)$, the $g_{\rho\eta\gamma}$
coupling constant is obtained from Eq.(\ref{eq21}) as: \bea
g_{\rho\eta\gamma} = (1.42 \pm 0.12) \label{eq22} \eea which is
very close to the sum rule prediction. %This result can be viewed as an
%indirect confirmation that the gluon component of the $\eta$ meson
%is absent or very small.

Finally we note that the coupling constant $g_{\omega\eta\gamma}$ can
be obtained from $g_{\rho\eta\gamma}$ with the help of the relation
$g_{\rho\eta\gamma}= 3 g_{\omega\eta\gamma}$

\newpage

\begin{figure}[h!]
$\left. \right.$ \vspace{-1cm}
\begin{center}
\includegraphics[width=12cm,angle=-90]{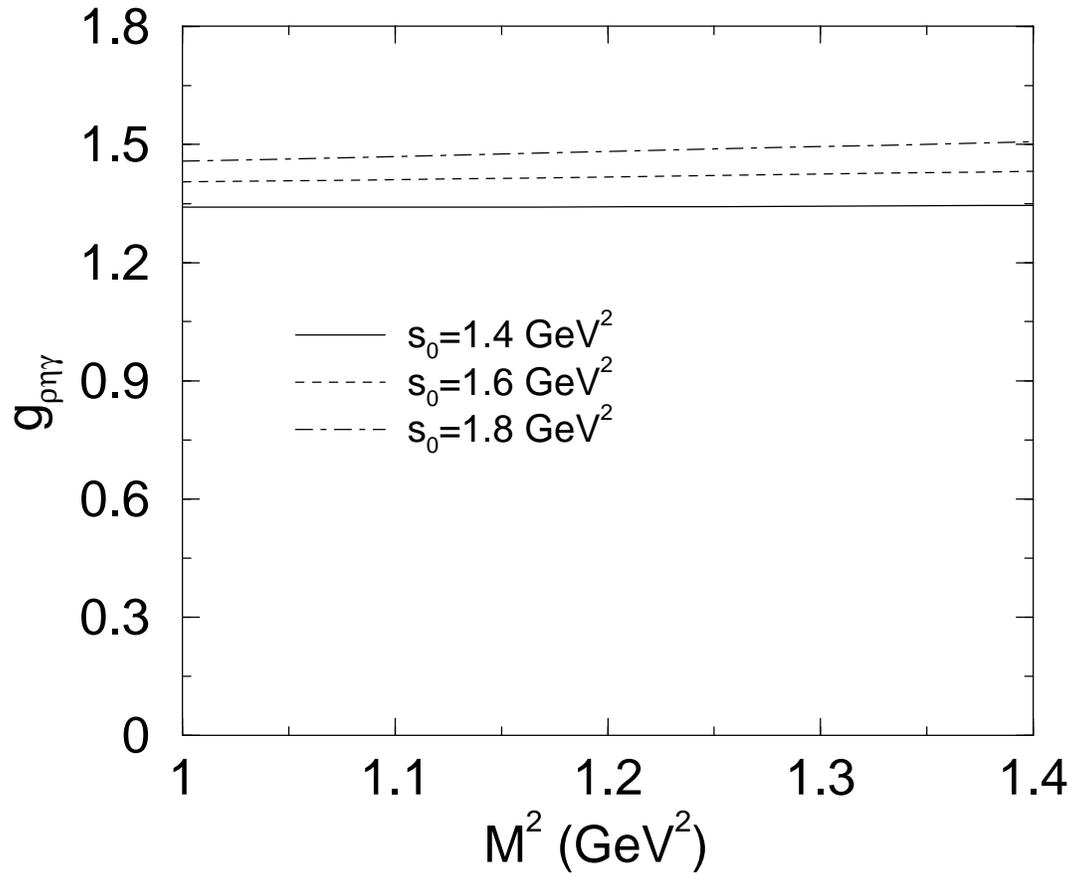}
\end{center}
\vspace{-1cm}
\caption{The dependence of the $g_{\rho\eta\gamma}$
coupling constant on the Borel parameter $M^2$ at three different
values of the continuum threshold $s_0= 1.4~GeV^2$, $s_0=1.6~GeV^2$
and $s_0=1.8~GeV^2$.}
\end{figure}

\end{document}